\documentclass[reqno,11pt]{amsart}
\usepackage[dvips]{changebar}
\usepackage[dvips]{color}
\usepackage{amssymb,amsthm}

\usepackage{a4}
\numberwithin{equation}{section}

\usepackage[mathscr]{eucal}

\usepackage{version}
\excludeversion{LONG}

\newcounter{mnotecount}[section]
\renewcommand{\themnotecount}{\thesection.\arabic{mnotecount}}

\newcounter{mymnotecount}[section]
\renewcommand{\themymnotecount}{\thesection.\arabic{mymnotecount}}
\newcommand{\mymnote}[1]
{\protect{\stepcounter{mymnotecount}}$^{\mbox{\footnotesize $
\bullet$\themnotecount}}$ \marginpar{
\raggedright\tiny\em $\!\!\!\!\!\!\,\bullet$\themymnotecount: #1} }
\renewcommand{\mymnote}[1]{}

\def\half{\frac{1}{2}}
\def\ben{\begin{equation}}
\def\een{\end{equation}}
\newcommand{\bea}{\begin{eqnarray}}
\newcommand{\eea}{\end{eqnarray}}

\theoremstyle{plain}
\newtheorem{thm}{Theorem}[section]

\begin{document}
\date{\today \ {\em File:\jobname{.tex}\,}}

\title{Gravitating Opposites Attract}

\author[R. Beig]{Robert Beig${}^{\dagger}$} \email{robert.beig@univie.ac.at}
\address{Gravitational Physics, Faculty of Physics, University of
Vienna, Boltzmanngasse 5, A-1090 Vienna, Austria}
\thanks{${}^{\dagger}$ Supported in part by Fonds zur F\"orderung der
Wissenschaftlichen Forschung project no. P20414-N16.}
\author[G.W.Gibbons]{Gary W. Gibbons} \email{G.W.Gibbons@damtp.cam.ac.uk}
\address{DAMTP, Centre for Mathematical Sciences, Cambridge University,
Wilberforce Road, Cambridge CW3 OBA, UK}
\author[R.M.Schoen]{Richard M. Schoen${}^\ddagger$} \email{schoen@math.stanford.edu}
\address{Department of Mathematics\\ Stanford University\\ Stanford, CA 94305}
\thanks{${}^\ddagger$Supported in part by NSF grant DMS-0604960}
\setcounter{tocdepth}{2}


\begin{abstract} Generalizing previous work by two of us,
we prove the non-existence of certain stationary configurations in
General Relativity  having a spatial reflection symmetry
across a non-compact surface
disjoint from the matter region. Our results cover cases such
that of two symmetrically arranged rotating bodies
with anti-aligned spins in $n+1$ ($n \geq
3$) dimensions, or two symmetrically arranged static bodies with
opposite charges in 3+1 dimensions. They also cover certain
symmetric configurations in (3+1)-dimensional gravity coupled to a
collection of
scalars and abelian vector fields, such as arise
in supergravity and Kaluza-Klein models. We also treat the  bosonic sector of simple supergravity in 4+1
dimensions.

\end{abstract}
\maketitle
\numberwithin{equation}{section}
\renewcommand{\theequation}{\thesection.\arabic{equation}}
\section{Introduction}
The title of this paper is an allusion to that  of one by  Aharonov,
Casher, Coleman, and Nussinov  \cite{ah} who give general arguments
why solitons in flat space, such as Yang-Mills monopoles, with the
opposite charges always attract. They showed, for example,  that for a
monopole-antimonopole configuration in Yang-Mills-Higgs theory,
one may always lower the energy by moving the monopole-anti-monopoles
together. This does not mean that the forces between them are necessarily
attractive, and indeed, Taubes \cite{Taubes} has shown
that there exist static monopole-antimonopole solutions.
However these solutions are unstable.

The forces considered by \cite{ah}
were classical, but in a subsequent  paper
 Kenneth and Klich
\cite{kk} showed that a similar property holds for quantum field-theoretic
  Casimir forces
\cite{kk}.
This was followed by a
paper of  Bachas \cite{Bachas} which considerably
 extended this result
and   related this  behaviour to the fundemental property of reflection
positivity in quantum field theory.

It is natural therefore to ask
whether the statement continues to hold in the presence of gravity,
according to  General Relativity. In the case of purely
gravitational forces between static bodies this question has been
addressed in a recent paper by two of us\cite{bs}. It was shown that
a bi-partite system of gravitating bodies \footnote{The word \lq \lq
bodies \rq \rq  should be taken to include ordinary bodies with
compactly supported energy momentum tensors and black holes with
regular event horizons} satisfying the conditions for which the
positive mass theorem holds, the two parts being separated by a
submanifold $S$ immersed in vacuum and which is totally geodesic
with respect to the the spatial metric, cannot  rest  in static
equilibrium.

The theorem proved in \cite{bs} holds in the case that the bodies
exert no long range forces other than gravity
upon each
other, in other words outside the bodies the Einstein vacuum
equations hold. We turn now to the case when the bodies carry
electric charges. In the static case we expect that there can be no
static equilibrium between oppositely charged bodies and a simple
modification of the method used in \cite{bs} allows us to prove this
({\bf Theorem 3.1}). Previous results, e.g \cite{g80}  assumed
axisymmetry. No such assumption is necessary for our result.

It is well known that rotating gravitating bodies, including black
holes, exert spin-spin forces \cite{w} which may be repulsive or
attractive depending upon the  orientations of their angular
momenta. For example an axisymmetric system of two bodies for which
their angular momenta are aligned in the {\sl same} direction along
the line of centres experiences a repulsion, while if the two
angular momenta are {\sl anti-aligned} they experience an attraction
\cite{w}. In the latter case (which we think of as opposites in a
sense to be made more precise later)  we do not expect an
equilibrium to be possible. The question of whether the
gravitational spin-spin interaction can ever overcome the
gravitational attraction is a long-standing one which has been
extensively studied for axisymmetric stationary metrics \cite{w94}
and recently been resolved \cite{nh} (for a discussion in 4+1
dimensions see \cite{hrzc}). As far as we are aware, there are no
rigorous results for stationary system which are not axisymmetric.
One of the principal aims of the present paper is to generalize the
argument given in \cite{bs} to prove in {\bf{Theorem 4.1}} without
assuming axisymmetry, that in the case of oppositely aligned spins
no such
equilibrium is possible.\\
To establish the result for rotating bodies it proves advantageous
to modify slightly the method of \cite{bs} by conformally rescaling
the spatial metric. Not only does this streamline some of the
calculations but it allows for an interesting physical
interpretation of  putative equilibria in terms of the balance of
stresses in the system, completely  analogous to the Newtonian
treatment of equilibria in terms of a gravitational stress tensor
$t^{U}_{ij}$ constructed  from the Newtonian potential $U$ and a
matter stress tensor $t^{\rm mat} _{ij}$. In the case of stationary
spacetimes there is a contribution to the stress tensor from  the
twist $\omega$ of the Killing vector. In the case of
Einstein-Maxwell theory, the matter stress tensor is constructed
from the electro-static and magneto-static potentials $\phi$ and
$\chi$.\\
This formalism allows an extension to cover much more complicated
matter systems. In particular it covers the systems discussed in
\cite{g82,bmg} which arise in supergravity, Kaluza-Klein and String
theories.

\section{Newtonian Considerations}
One of the first things we are taught in physics is that like
charges repel and unlike charges attract. By \lq \lq charges \rq \rq
is usually meant electric charges and for  example, one could
consider the force between a pair of apparently identical sources.
If the two sources have the same electric  charge, there will be a
repulsive force and if they have the opposite electric charge there
will be an equal but opposite attraction. The same holds for
magnetism, and indeed it seems that historically  what is usually
called Coulomb's law was first established for  magnetic poles by
John Michell \cite{wh}, inventor of the torsion balance  subsequently used
by Cavendish to measure Newton's constant $G$, and of the concept of
a black hole. Michell would have been aware that because  gravity is
always attractive an equilibrium  between two equal masses with
opposite charges is never possible. If the charges are the same,
then of course  by special choice of the charges and masses an
equilibrium may be possible. More generally one may consider a
bi-partite system in which two disjoint  sets of charges and masses,
or a continuous distribution of matter and electric  are
symmetrically disposed with respect to a  plane   $S$ such that the
charges on one side are the opposite of the charges on the other
side. The net gravitational force between them is clearly attractive
\footnote{provided of course that the gravitational masses are taken
to be positive, which is not mandatory in Newtonian theory
\cite{Foppl}}
and the net electric force between them will also be
attractive and again no equilibrium is possible. A formal
justification of this might proceed by considering the electrostatic
and Newtonian gravitational stress tensors
\begin{eqnarray}
t^{\phi}_{ij} &=& (D_i \phi)  (D_ j \phi) - \frac{1}{2}
\delta_{ij} (D \phi)^2 \label{estress}\\
t^{U}_{ij}   &=& - \frac{1}{4 \pi G}\, (D_i U)(D_ j U) + \frac{1}{8
\pi G}\, \delta_{ij} (D U)^2 \label{gstress}
\end{eqnarray}
where $\phi$ and $U$ are the electrostatic and gravitational
potentials.

Since the electric field lines which start on the positive charges
must end on the negative charges, and symmetry dictates that
$E_i=-\partial_i \phi $ is orthogonal to $S$
 there is a net electric flux crossing
$S$ and therefore a net attractive electric
 force
\begin{equation}
F_i^{\phi}  = \int _{S}  t^{\phi} _{ij} \,d \sigma _j
\end{equation}
The gravitational potential must by contrast be constant on $S$
but because of the opposite sign in (\ref{gstress}) compared with
(\ref{estress}) there is again a net attractive force between the
systems and no equilibrium is possible.

A general necessary condition for equilibrium in flat spacetime
\begin{equation}
\int _{S}  t _{ij} \,d \sigma _j=0\,, \label{stress}
\end{equation}
where $t_{ij}$ is the total stress tensor. In some cases it may be
used  to rule out the existence of certain  static equilibria. For
example one can show that (\ref{stress}) can not be satisfied on any
plane separating the bodies. Conversely, where such a plane does not
exist, one can in fact construct 2-body configurations \cite{bs8},
and analogues have been found in GR in \cite{as}. These equilibrium
configurations will in general not be stable. Let us remark that
nowhere in this paper do we touch the issue of
stability.
In what follows
we shall generalize condition (\ref{stress}) to incorporate the
effects of General Relativity.

In flat space the necessary condition (\ref{stress})
must   hold in particular
for the Yang-Mills-Higgs equations considered
in \cite{ah,Taubes}. If the Higgs potential vanishes
and the fields satisfy the first order Bogomolny
equations, then the total stress tensor $t_{ij}$ is known to vanish
pointwise. This is consistent with the existence of the well known static
multi-monopole  solutions  of the first order equations.
We shall show shortly that a similar statement holds
for static charged multi-black hole solutions in general relativity.
Taubes's monopole-antimonopole solutions  however
satisfy the second order equations but not the first order equations.
It is not obvious to us whether (\ref{stress}) is satisfied
because the stresses vanish pointwise on $S$, or
because a cancellation in the integral.

Of course  the notion of force does not make
much sense in General Relativity, but the notion of a stress
tensor for the electromagnetic field certainly does and, as
we shall see shortly, as  does, if the system is static,
the idea a gravitational stress tensor. This will allow us to
implement the ideas above in a general relativistic
context and to extend the recent proof \cite{bs}
on the non-existence of  static n-body configurations
in pure gravity to the electrostatic case.

Something which was not envisaged by Newton, although it
has a correspondence in Newtonian theory is the frame-dragging
forces  exerted by rotating bodies. We shall show, in effect by introducing
a suitable stress tensor for magneto-gravitational forces,
how the results of \cite{bs} can be extended to stationary bodies.

For a general discussion of the fact that forces mediated by fields of
even(odd) spin are attractive(repulsive) see \cite{Deser}.

\section{Electrostatic case}

We begin, for the sake of simplicity, by giving the simplest
generalization of the argument of \cite{bs}.

If we  write the static metric as
\begin{equation}
ds ^2 = -V^2 dt ^2 + g_{ij} dx^i dx^j\,\qquad A= \phi\,
dt,\,\,\,\,\kappa = 8 \pi G
\end{equation}
\bea R_{ij}&=& \frac{1}{V} D_i D_j V - \frac{\kappa}{V^2}
 \Bigl[(D_i \phi)(D_j \phi) -\half g_{ij} (D \phi)^2 \Bigr]\,,
 \label{one}\\
V\,\Delta V&=& \frac{\kappa}{2} (D \phi)^2  \,,\\
V \Delta \phi &=& g^{ij} (D_i \phi)(D_j V) \,. \eea

As a check, note that we obtain the standard Hamiltonian constraint
\ben R=
\frac{\kappa}{V^2}(D \phi)^2\,, \een as expected

Following \cite{bs} we assume that the space of orbits, $N$,  of the
timelike Killing field  admits a totally geodesic
surface $S$ with Gauss-curvature $K$, then \ben R_\alpha{}^\alpha=2K
+ R_{nn}\,, \een where the first term on the left is  the tangential
trace with respect to the metric on $S$ and $R_{nn}$ is defined by
$R_{nn} = R_{ij}n^i n^j$, with $n^i$ being the unit normal of $S$.
Taking the tangential trace of (\ref{one}) gives \ben
R_\alpha{}^\alpha = \frac{1}{V} \Delta_S V +
 \frac{\kappa}{V^2} (D_n \phi )^2
\een

Thus we arrive at \ben K= \frac{1}{V} \Delta_S V + \frac{\kappa}{2
V^2} \Bigl [(D_n \phi )^2 - (D_\alpha \phi)(D^\alpha \phi) \Bigr]\,,
\label{formula} \een\

If we assume that $S$ is an iso-potential of $\phi$ we have $
D_{\alpha} \phi   =  0$ and we may proceed as in \cite{bs}.

\subsection{Unlike charges}
Our assumptions would hold if we had a static system of charged
bodies invariant under an isometric action of ${\mathbb Z}_2 $ which
stabilizes $S$ pointwise  and under  which the electric field is
{\sl odd} \ben {\mathbb Z}_2 : D \phi \rightarrow - D \phi \,. \een
This is the analogue  of the situation considered in \cite{ah} for
solitons and in \cite{kk,Bachas} for Casimir forces. Thus we obtain
the
\begin{thm} Assume that $(N,g)$ is static electrovacuum outside a compact set and has $R\geq 0$ everywhere.
Suppose there is a properly embedded, noncompact, totally geodesic
surface $S$ such that $g$ is static electrovacuum in a neighborhood
of $S$ and that the pull-back-to-$S$ of $D_i \phi$ is zero. It
follows that $(N,g)$ is isometric to the Euclidean space ${\mathbb
R}^3$ and $(M,ds^2) = (\mathbb{R} \times N, - V^2 dt^2 + g)$ is
Minkowski space.
\end{thm}
For the proof we refer the reader to the generalization of this
result to Einstein-Maxwell-Dilaton theory in Sect.5.


\subsection{Like Charges}

In this case we assume  that  electric field is {\sl even} under the
${\mathbb Z}_2$ action \ben {\mathbb Z}_2 : D \phi \rightarrow D \phi \,.
\een We also have, in both even and odd  cases \ben D_n V=0\,, \een
Now we know of one example, the
Majumdar-Papapetrou solutions \cite{syn} for which equilibrium is
possible. This is if \bea
D \phi &=& \pm \sqrt{\frac{2}{\kappa}} \, D V\,.\\
g_{ij} &=& V^{-2} \delta_{ij} \,. \eea and in fact $V^{-1}$ is
harmonic wrt to the flat metric $\delta_{ij}$. In that  case $S$ is
conformally flat and hence  $K \ne 0$. However \ben \int _S K = \int
_S \frac{1}{V^2} \Bigl[(D_\alpha V)(D^\alpha V) + \frac{\kappa}{2}
(D_n \phi)^2 -  \frac{\kappa}{2} (D_\alpha \phi)(D^\alpha \phi)
\Bigr]=0\,, \een as required.

A complete treatment of the   necessary and sufficient conditions
for equilibrium of charged gravitating objects with the same sign in
General Relativity is not  yet completely available (see
\cite{Bon}).  The most recent results in the axisymmetric case can
be found in \cite{Man}.

\section{Stationary case in $n+1$ dimensions}
We assume $(M, ds ^2)$ has a timelike Killing vector $\xi^\mu$ with
complete orbits. Then $ds^2$ can be written as
\begin{equation}\label{metric}
ds^2  = - e^{2 U} (dt + \psi_i dx^i)^2 + e^{- \frac{2 U}{n-2}}\,
h_{ij} dx^i dx^j
\end{equation}
with $h$ being a Riemannian metric on $N$, the quotient space under the action of $\xi = \partial_t$.
We refer to $U$ as the gravitational potential and define $1/2$ the curvature of the Sagnac connection $\psi$ by
\begin{equation}\label{sagnac}
\omega = \frac{1}{2}\, d \psi,\hspace{1.2cm}\omega_{ij} = \partial_{[i} \psi_{j]}
\end{equation}
We only need the gravitational part of the action. There is the
identity
\begin{equation}\label{scalar}
R \sqrt{-g} = \left[\frac{2}{n-2}\,\, \Delta_h U + \mathcal{R} -
\frac{n-1}{n-2}\, (DU)^2 +
e^{2U\frac{n-1}{n-2}}\,\omega_{kl}\omega^{kl}\right] \sqrt{h}\hspace{0.2cm},
\end{equation}
where $\mathcal{R}$ is the Ricci scalar of $h_{ij}$. Hence, modulo a
surface term, the reduced gravitational action is given by
\begin{equation}\label{action}
S = \int \left[\mathcal{R} - \frac{n-1}{n-2}\, (DU)^2 +
e^{2U\frac{n-1}{n-2}}\,\omega_{kl}\omega^{kl}\right] \sqrt{h}\, d^3
x
\end{equation}
and the vacuum field equations turn out to be ($\mathcal{G}_{ij} =
\mathcal{R}_{ij} - \frac{1}{2} h_{ij} \mathcal{R}$)
\begin{subequations}\label{field}
\begin{align}
\mathcal{G}_{ij} + 8 \pi G\,(t^U_{ij} + t^\omega_{ij})&= 0\\
\Delta_h U + \,e^{2 U \frac{n-1}{n-2}}\,\omega_{kl}\omega^{kl}&=0\\
D^j\left(e^{2 U \frac{n-1}{n-2}} \,\omega_{ij}\right)& = 0
\end{align}
\end{subequations}
In (\ref{field}a) we are using the definitions
\begin{equation}\label{def1}
8 \pi G \,t^U_{ij} = \frac{n-1}{n-2}\, [- (D_i U) (D_j U) +
\frac{1}{2}h_{ij}(DU)^2]
\end{equation}
and
\begin{equation}\label{def2}
8 \pi G \,t^\omega_{ij} = 2 \,e^{2 U \frac{n-1}{n-2}}\,[\,
\omega_{ik}\omega_{j}{}^k - \frac{1}{4} \,h_{ij}\,
\omega_{kl}\omega^{kl}]
\end{equation}
(Using $d \omega = 0$, Eq.(\ref{field}c) follows from Eq.'s (\ref{field}a,b).) We consider asymptotically flat solutions of (\ref{field}).
They can be shown (see e.g. \cite{mp})
to have the form
\begin{subequations}\label{as}
\begin{align}
h_{ij}&=\delta_{ij} + o^\infty\left(\frac{1}{r^{n-2}}\right)\\
U &= - \frac{8 \pi}{(n-1) A_{n-1}}\frac{m}{r^{n-2}} + o^\infty\left(\frac{1}{r^{n-2}}\right)\\
\omega_{ij} &= - \frac{4 \pi}{A_{n-1}}\frac{L_{ij} + n
\,a_{[i}L_{j]k}a^k}{r^n} + o^\infty\left(\frac{1}{r^n}\right)
\end{align}
\end{subequations}
Here $a^i = \frac{x^i}{r}$ and $A_n$ is the area of $\mathbb{S}^n$.
Furthermore the constants $m$ and $L_{ij}=L_{[ij]}$ are respectively
the ADM mass and the (mass-centered) spin tensor of the
configuration. When $m=0$, the ADM energy of the initial data set
induced on $t=\mathrm{const}$ is also zero. Then, when the source
satisfies the dominant energy condition, the positive energy theorem
implies that spacetime is flat\footnote{When $n=3$ one can also
allow for the presence of horizons, see \cite{ghp},\cite{bc}.}.
When there is more than one asymptotically
flat end, all the statements above hold separately
w.r. to any such end.\\
Suppose $(N,h)$ admits a reflection symmetry $\Psi$ across an
$(n-1)$ - dimensional surface $S$, which is disjoint from the matter
region. There are two ways this can be lifted to an isometry of
$(M,g)$. One is that if, in addition to $\Psi^\star h = h$, there
holds $\Psi^\star U = U$ and $\Psi^\star \omega = \omega$. Then,
provided that $N$ is simply connected,
there is a reflection of $(M,g)$ which preserves $\xi$.
(This reflection preserves the timelike hypersurface $\Sigma$
pointwise consisting of the orbits of $S$ under the flow of $\xi$.)
The other possibility is that if $\Psi^\star U = U$ and $\Psi^\star
\omega = - \omega$, in which case there is a reflection of $(M,g)$
which maps $\xi$ to $- \xi$. In both cases we have that $D_n U|_S =
0$, where $n^i$ is a normal of $S$ and that $\omega_{ij} n^j|_S = 0$
in the first case and that the pull-back-to-$S$ of $\omega$ be zero
in the
second case. \\
When the totally geodesic hypersurface $S$ is
properly embedded
, non-compact and closed, by a straightforward extension of
Propositions 2.1. and 3.1. in \cite{bs}, $S$ can be shown to be
asymptotic to a disjoint union of a finite number of hyperplanes at
infinity.
Let some such
asymptotic end of $S$ be given by $x^n = 0$. It then follows from
(\ref{as}) that the spin tensor
has to satisfy $L_{\alpha n}=0$ in the first case and $L_{\alpha \beta} = 0$ in the second, where $\alpha,\beta = 1,...n-1$.\\
Now the stress energy tensor of the matter source can be viewed as a
superposition of two stress energy tensors $T_{\mu\nu}^{\mathrm{mat}}$ and
$T_{\mu\nu}^{'\,\mathrm{mat}}$ which are images of each other under the map $\Psi$
(i.e. $(\Psi^\star T)_{ij}^{\mathrm{mat}}=T_{ij}^{'\,\mathrm{mat}},\,
\Psi^\star \rho = \rho',\,
(\Psi^\star j)_i = \pm \,j'_i$ in obvious notation). Then,
heuristically, we have that $L_{ij} = L_{ij} + L'_{ij}$ with $L, L'$
being the 'spins of the individual configurations'. Thus there holds
$L_{\alpha \beta}=L'_{\alpha \beta}$ and $L_{\alpha n} = - L'_{\alpha n}$ in the first case and
$L_{\alpha \beta}=-L'_{\alpha \beta }$ and $L_{\alpha n} = L'_{\alpha n}$ in the second case. We now
argue that these two cases correspond to a repulsive (resp.
attractive) spin-spin force between the two configurations. To see
this we use, following \cite{w},  the Mathisson-Fock-Papapetrou
expression for the force $F_i$ on a particle at rest, i.e.

{\begin{equation}\label{math}
F_i = -
\frac{1}{2}L_{\rm test}^{jk}R_{jki0}\hspace{2cm}\mathrm{large}\hspace{0.3cm}r
\end{equation}
Here $L_{\rm test} ^{ij}$ is the spin tensor of a test particle. Now from the
Killing identity
\begin{equation}\label{kill}
\nabla_\mu \nabla_\nu \xi_\lambda = - R_{\nu \lambda \mu \rho}\, \xi^\rho
\end{equation}
applied to the Killing vector $\xi^\mu \partial_\mu = \partial_t$ we
find, using (\ref{metric},\ref{sagnac}), that
\begin{equation}\label{ident}
\partial_i \omega_{jk} = R_{jki0}
\end{equation}
to leading order. It follows that the spin-spin force is attractive
(resp. repulsive) to leading order in $1/r$, whenever the quantity
$\omega_{ij}{ L_{\rm test}}  ^{ij}$ is negative (resp. positive).
This, in turn, is
the same as the quantity
\begin{equation}\label{int}
L_{ij}^{\rm source} L_{\rm test} ^{ij} - n L_{ij}^{\rm source}L_{\rm test}^i{}_k n^j n^k = L_{\alpha \beta}^{\rm source} L_{\rm test} ^{\alpha \beta} -
(n-2)L_{\alpha n}^{\rm source} {L_{\rm test}}  ^\alpha{}_n
\end{equation}
being negative (resp. positive), where $n^i$ is the unit vector
pointing from the source to the test particle. So, indeed, this
is consistent with the  force between rotating black holes being
 attractive when $L_{\alpha \beta}= - L'_{\alpha \beta}$ and $L_{\alpha n} = L'_{\alpha n}$
and repulsive when $L_{\alpha \beta}= L'_{\alpha \beta}$ and $L_{\alpha n} = - L'_{\alpha n}$.\\
For example in 3+1 dimensions, at large separation   $r = |{\bf
r}|$, the mutual potential energy of two spinning bodies with
angular momentum ${\bf{J}}$ and ${\bf{J}}'$ ($J_i = \epsilon_{ijk} S^{jk}$) is given by \cite{w}
\begin{equation}
\frac{G}{c^2 r^3} \Bigr[  3  ( {\bf n}\cdot {\bf J}) (  {\bf n}
\cdot{\bf J}')   - r^2 ( {\bf J} \cdot {\bf J}')   \Bigl] \,.
\end{equation}
This gives an  attractive force  if ${\bf J}= -{{\bf J}'}$  and a
repulsive force  if ${\bf J} =+ {\bf J}'$. Our results are
consistent with this. They are also consistent with the behavior of
explicit exact  {\it double Kerr} solutions obtained using
solution-generating techniques. These exhibit conical singularities
along the axis between the two sources which is interpreted as a
strut or rod in tension which holds the two black holes apart. For a
recent detailed discussions see \cite{hr,nh,chr}.\\
We now state as the main result of this section the
\begin{thm} Let $(N,h)$ have a hypersurface $S$ disjoint from
the matter region, which is non-compact, closed and totally geodesic
w.r. to the unrescaled metric $e^{- \frac{2 U}{n-2}} h_{ij}$. (Our
previous remarks on possible 'liftings' of isometries from $(N,h)$
to $(M,ds^2)$ are equally valid when applied to $(N,e^{- \frac{2
U}{n-2}} h)$.) Suppose in addition that the pull-back-to-$S$ of
$\omega$ is zero. Then spacetime is flat.
\end{thm}
{\bf{Remarks:}} By the discussion above, this result covers the
'good' case, in which there is no chance for the spin-spin force to
balance the
gravitational attraction. Note also that we do not assume $D_n U|_S = 0$.\\
{\bf{Proof:}} Contracting (\ref{field}a) with $n^i n^j$, with $n^i$
being the unit normal to $S$ in $(N,h)$ we find
\begin{equation}\label{nn}
\mathcal{G}_{nn} = \frac{n-1}{n-2}\left[\frac{1}{2}\,(D_n U)^2 -
\frac{1}{2}\,(D_\alpha U)(D^\alpha U)\right] - \,e^{2U
\frac{n-1}{n-2}} \,\omega_{\alpha n} \omega^\alpha{}_n\,\,.
\end{equation}
Here $D_\alpha$ is the intrinsic derivative on $S$. In (\ref{nn}) we
have used that $\omega_{\alpha \beta}$ is zero. Now, from the Gauss
equation,
\begin{equation}\label{gauss}
\mathscr{R} = - 2\, \mathcal{G}_{nn}  + (\mathrm{tr}\, k)^2 -
\mathrm{tr}(k^2)\,,
\end{equation}
where $\mathscr{R}$ is the Ricci scalar of $S$ and $k$ its extrinsic
curvature. Since $S$ is totally geodesic w.r. to $e^{- \frac{2
U}{n-2}} h$, we have that
\begin{equation}\label{bar}
k_{\alpha \beta} = \frac{1}{n-2}\,\, q_{\alpha \beta} \,D_n U
\end{equation}
where $q$ is the intrinsic metric on $S$ induced from $h$. Inserting
(\ref{bar},\ref{nn}) into Eq.(\ref{gauss}), the terms involving $D_n
U$ cancel so that finally
\begin{equation}\label{pos} \mathscr{R} = \frac{n-1}{n-2}\, (D_\alpha
U)(D^\alpha U) + 2 \,e^{2U \frac{n-1}{n-2}}\,\omega_{\alpha n}\omega^\alpha {}_n\;.
\end{equation}
In particular $\mathscr{R}$ is non-negative. By virtue of the
'asymptotically planar' nature of $S$ (see \cite{bs} for details)
the metric $q$ on $S$ tends to the Euclidean metric on $S$ as fast
as $h$ tends to the flat metric on $N$
, so $S$ has zero ADM mass. Now for $n>3$ we can apply
the positive energy theorem
\footnote{When $n=4$ this is the standard positive energy theorem in 3
dimensions \cite{sy}. For the positive energy theorem in general dimensions see \cite{s}.}
to $S$
yielding that $S$ is flat $\mathbb{R}^{n-1}$, in particular
$\mathscr{R}$ is zero, whence $U|_S$ is zero. The same conclusions
are reached for $n=3$, by using the Gauss-Bonnet theorem.
Consequently $m=0$ and spacetime is flat, again by the general positive energy theorem.
\section{A General Sigma-Model formalism}
In \cite{bmg} a general formalism was developed to cover stationary solutions
of Einstein's equations  in 3+1 dimensions coupled to  the matter sectors of various
supergravity or Kaluza-Klein type theories theories.
The models typically contain $n_s$ scalar fields and $n_v$ abelian
vector fields.
The same metric ansatz was made, and after some  work
all vector fields were swapped for $n_v$ generalised electrostatic potentials
and $n_v$  magneto-static potentials. All such fields
could be combined into a collection of  scalar fields  $\Phi^A$,
where $A=1,2 \dots, 2+n_s + 2 n_v $. The fields $\Phi^A$ include
the Newtonian potential $U$ and  the Sagnac curvature $\omega$, whose
precise definition depends on the matter fields considered
and may be regarded as providing a map into some target space
$M_\Phi$ with, a metric $G_{AB}(\Phi^C) $ with signature $(2+n_s,2 n_v)$.
In many cases $M_\Phi, G_{AB} (\Phi)$ is a pseudo-riemannian
symmetric space $G/H$. The effective Lagrangian is
\begin{equation}\label{effective}
\int \left[\mathcal{R}- 2 h^{ij}
 G_{AB} (\partial_i \Phi ^A)(\partial _j \Phi ^B)
\right] \sqrt{h}\, d^3 x
\end{equation}
Thus the equations of motion reduce to the statement that
the map $\Phi^A(x)$  be harmonic and that
\begin{equation}\label{effective1}
\mathcal{R}_{ij}-\frac{1}{2} \mathcal{R} h_{ij} = 8 \pi G t^\Phi_{ij}
\end{equation}
where
\begin{equation}\label{effective2}
4 \pi G t^\Phi_{ij} = G_{AB}(\partial_i \Phi ^A)(\partial_j \Phi ^B) -\frac{1}{2}
h_{ij}  h^{kl}
 G_{AB} (\partial _k \Phi^A)(\partial _l \Phi ^B)\,.
\end{equation}
Like in (\ref{nn}) we see that for any chance of an equilibrium the quantity $t^\Phi_{nn}$ given by
\begin{equation}\label{effective3}
- 8 \pi G t^\Phi_{nn} = G_{AB}[(D_\alpha \Phi^A)(D^\alpha \Phi^B) -
(D_n \Phi)^2]
\end{equation}
should have some positive contributions other than the $(D_n U)^2$ -
term, which is zero anyway in the presence of a spacetime reflection
symmetry across $S$. Thus there should be a {\it{pressure}} rather
than a {\it{tension}} across the surface $S$.

In fact we can regard the $U$ contribution
as a purely gravitational contribution to the effective stress tensor.
The formula
\begin{equation}\label{effective4}
\int _S \mathscr{R}= 2 \int_S K = 0\,
\end{equation}
is the statement that the total stresses (gravitational and matter)  on the
surface $S$ should vanish. As noticed by Maxwell, gravitational
field lines are in compression and exert a tension
transverse to the field lines, the opposite behaviour
to that of the  electric field lines.

If $S$ were  not totally geodesic, it seems that
one gets a contribution quadratic in the  extrinsic curvature of $S$ which is like a
{\it bending energy}.

\subsection{Maximum Tension}
If we don't require that the system be asymptotically flat, but
merely that $S$ have the topology of ${\mathbb R}^2$, we have (for an
outline of the proof and references see \cite{g93})
\begin{equation}
\int _S K < 2 \pi
\end{equation}
which yields and upper bound on the tension \cite{g02}
\begin{equation}
-\int t_{ij}n^in^j \le \frac{1}{4G} \,.
\end{equation}


\subsection{Electrostatic Example}
For this the target space is $SO(2,1)/SO(2)$,
\begin{equation}
G_{AB}d \Phi^A d\Phi^B= dU^2 - e^{-2U} d \phi ^2
\end{equation}
and
\begin{equation}
8 \pi G t^\Phi_{ij} = 2\, [(D_i U) (D_j U) -
\frac{1}{2}h_{ij}(DU)^2] - 8 \pi G e^{- 2 U}[(D_i \phi)(D_j \phi) - \frac{1}{2} h_{ij} (D \phi)^2]
\end{equation}
Suppose we have a static system of charged bodies with a reflection isometry across $S$ under which
$d \phi$ is odd, i.e. we have unlike charges. This is the analogue  of the situation considered in \cite{ah}
for solitons and in \cite{kk} for Casimir forces. Then the nonexistence-argument in the Theorem in Sect.2 goes through unchanged. The argument would also go through for Einstein-Maxwell-Dilaton theory.

In the case of like charges, as mentioned in Sect.(3.2), there are in fact equilibrium solutions,
namely in the Majumdar-Papapetrou case, where $dU= \pm \sqrt{4 \pi G}\, e^{-U}\,d \phi$ (whence $t^\phi_{ij}=0$),
$h_{ij}$ is flat and $D^i(e^{- U} D_i \phi) = 0$. These solutions describe black holes of equal mass and charge where the gravitational attraction is exactly balanced by electrostatic repulsion.

In the Majumdar-Papapetrou
case a  boost inside $SO(2,1)$ gives the
Harrison transformation and one may boost the well-known Israel-Khan
multi-black hole solution \cite{Israel} to give a
multi-charged solution which however
is non-singular along the axis only in the
Majumdar-Papapetrou limit \cite{g80}.
\section{4+1 dimensions}
Many recent interesting applications of
black holes, black rings etc have involved simple supergravity in
4+1 spacetime dimensions. The bosonic sector of this theory is
essentially Einstein-Maxwell theory supplemented with a crucial
Cherm-Simons term whose coefficient is dictated by supersymmetry.
The equations for stationary metrics may be derived from an action
principle given in \cite{lmp}.

The Lagrangian 5-form  for the five-dimensional theory is

\begin{equation}
R\eta  -  \frac{1}{2} {* F}\wedge  F + \frac{1}{\sqrt{27}}\,  F\wedge  F\wedge  A\,,
\end{equation}
where $\eta$ is the volume 5-form and $ F=d A$, and
$ A$ is the electromagnetic  potential.\footnote{
This implies that the graviphoton equation of motion is
$\nabla_\nu  F^{\mu\nu} + 1/(4\sqrt3) \epsilon^{\mu \alpha\beta\gamma\delta} F_{\alpha\beta}  F_{\gamma\delta}=0$.}

One makes the ansatz
\begin{eqnarray}
ds^2  &=& - e^{-2\phi}\, (dt + \omega )^2 + e^{\phi} \gamma_{ij} dx^i dx ^j\,\\
 A & =& \tilde A   + \sqrt3 \chi (dt+  \omega)\,.
\end{eqnarray}

The Lagrangian 4-form in 4 dimensions  is
\begin{equation}\label{4action}
 \tilde R\, \tilde \eta   -\frac{3}{2} {*_\gamma d\phi}\wedge d\phi +
           \frac{3}{2}e^{2\phi} { * _\gamma d\chi}\wedge d\chi +
       \frac{1}{2} e^{-3\phi}{*_\gamma  F }\wedge  F
 - \frac{1}{2} e^{-\phi} {*_\gamma \tilde F}\wedge \tilde F
  + \chi d \tilde A\wedge d \tilde A\,,
\end{equation}
where
\begin{equation}
F = d \omega \,,\qquad \tilde F= d \tilde A + \sqrt3 \chi d \omega
\,.
\end{equation}
Setting $\chi=0=\tilde A $ we obtain the purely
gravitational theory.
$\phi$ is the Newtonian potential and $F$ the gravito-magnetic field.
The electrostatic potential is  $\chi$ and magnetic field is $\tilde
F$. If one evaluates the effective stress tensor one sees that the last
term in (\ref{4action}) does not contribute since it is ``\lq \lq
topological \rq \rq, i.e. it  may be written
without any metric. Thus the stress tensors
for $\chi$ and $\tilde F$ are identical in structure but opposite in sign
to those for $\phi$ and  $F$.

It is the structure of the term involving $F_{ij}$ which
is of principal interest.
Let us define
\begin{equation}
t^{\rm spin} _{ij} =  F_ {ik} F_{jl} \gamma ^{kl} - \frac{1}{4}
\gamma _{ij} F_{kl} F_{rs} \gamma ^{kr} \gamma ^{ls} \,.
\end{equation}
The  scalar  curvature of the totally geodesic surface $S$
contains a term proportional to
\begin{equation}
+2t^{\rm spin} _{nn} \,.
\end{equation}
Defining
\begin{equation}
\mathcal{E}_{\alpha} = F_{n \alpha}\,, \qquad \mathcal{B}_3 = F_{12}
\,,\qquad {\rm etc.}
\end{equation}
one has
\begin{equation}
t^{\rm spin} _{nn}= \frac{1}{2}  \sum _{\alpha = 1,2 ,3} \bigl(
\mathcal{E}_\alpha \mathcal{E}^\alpha - \mathcal{B}_\alpha
\mathcal{B}^\alpha
               \bigr )
\end{equation}

This is positive or negative depending  upon whether takes the case in which
$F_{ij} $ is  even or odd under reflection in $S$.

The resultant pattern of attractions and repulsions
is presumably consistent with the detailed models discussed in
 \cite{hrzc}.

\section{Conclusion}
In this paper, extending  previous work two of us \cite{bs},
we have studied the
stationary equilibrium of a system of gravitating bodies
in General Relativity, possibility coupled to  to one or more electromagnetic
and scalar fields, which   is  reflection symmetric about a surface which
does not intersect the sources. In the case that the reflection symmetry
acts so that  that the bodies  on one side have the opposite charges
and angular momenta to those on the other side,
we were  able to exclude such a possibility.

Our  principal innovation was  to make use of a suitable
conformal rescaling of the spatial metric which allowed  us to express
the necessary condition for
equilibrium in terms of a stress tensor for the gravitational,
gravito-magnetic fields and any possible scalars or vector fields.
The necessary condition resembles closely the elementary requirement
for any equilibrium in flat spacetime, that the integrated total  stress across
any surface must vanish.
We have applied our methods to stationary (n+1) dimensional
vacuum spacetimes for arbitrary $n\ge 3$.
We have also applied them to Einstein Maxwell theory
in $3+1$ dimensions and supergravity theories in $4+1$ dimensions as well
as to various theories in $3+1$ which arise
in super gravity and Kaluza-Klein theories.

An interesting question for further study is whether our methods may be
applied
to the equilibrium of various extended objects (\lq \lq p-branes \rq \rq )
which arise in String/M-Theory.

A bigger challenge would be to extend the methods to encompass
a cosmological constant, either positive or negative.
In the case of negative cosmological constant this might be possible
using the various extensions of the positive mass theorem to
asymptotically anti-de-Sitter  spacetimes.
In the case of positive cosmological constant it is less obvious how to
proceed.

\subsection*{Acknowledgement} One of us (G.W.Gibbons) would like to acknowledge conversations with Carlos Herdeiro.


\begin{thebibliography}{XXX}


\bibitem[ACS] {ah} Aharonov, Y., Casher, A., Coleman, R.S., and Nussinov, S. (1992) {\it{Why opposites
attract}}, Phys. Rev.D {\bf{46}}, 1877-1878



\bibitem[AS] {as} Andersson, L. and Schmidt., B.G. (2009) {\it{Static self-gravitating many-body systems in Einstein
gravity}}, [arXiv:0905.1243 [gr-qc]].

\bibitem[BA]{Bachas} Bachas, C.P. (2007) {\it Comment on the sign of the
Casimir force} J.Phys.{\bf A40} 9089-9096 [arXiv:quant-ph/0611082]

\bibitem[BC] {bc} Bartnik, R.A. and Chrusciel, P.T., (2005) {\it{Boundary value problems
for Dirac-type equations}} J. Reine Angew. Math.{\bf{579}}, 13-73
[arXiv:math/0307278]

\bibitem[BS]{bs8} Beig, R. and Schmidt, B. G. (2008) {\it{Celestial mechanics of elastic bodies}},
Math. Z. {\bf 258} 381--394, [arXiv:gr-qc/0612189]

\bibitem[BSc] {bs} Beig, R. and Schoen, R.M. (2009) {\it{On Static n-Body Configurations in Relativity}},
Class.Quant.Grav.{\bf{26}} 075014 (7pp), [arXiv:0811.1727 [gr-qc]]

\bibitem[B]{Bon} Bonnor, W. B.(1993) ,
{\it The equilibrium of a charged test particle in the
field of a spherical charged mass in general relativity},
 Class. Quantum Grav. {\bf 10}  2077-2082

\bibitem[BMG] {bmg} Breitenlohner, P., Maison, D.,  and Gibbons, G.W. (1988) {\it{Four-Dimensional Black
Holes from Kaluza-Klein Theories}}, Commun. Math. Phys. {\bf{120}},
295-333.

\bibitem[CHR] {chr} Costa, M.S., Herdeiro, C.A.R., and Rebelo, C.
(2009) {\it{Dynamical and Thermodynamical Aspects of Interacting
Kerr Black Holes}}, [arXiv:0903.0264[gr-qc]]

\bibitem[D] {Deser} Deser, S. (2005) {\it How Special Relativity Determines the Signs of the Nonrelativistic, Coulomb and Newtonian Forces} Am. J. Phys.{\bf{73}}, 752 [arXiv:gr-qc/0411026]

\bibitem[F]{Foppl} F\"oppl, A. (1897)  {\it  \"Uber eine
Erweiterung des Gravitationsgesetzes}, Sitzungsber. d. M\"unch. Akad
{\bf 6}, see also F\"oppl, A.  (1910) {\it Vorlesungen \"uber
technische Mechanik, VI, Erster Abschnitt, Die relative Bewegung},
Teubner
\bibitem[GH] {gh} Gibbons, G.W. and Hawking, S.W. (1979) {\it{Classification Of Gravitational Instanton
Symmetries}}, Commun. Math. Phys.{\bf{66}} 291-310.

\bibitem[GHP] {ghp} Gibbons, G.W., Hawking, S.W., Horowitz, G.T.,
and Perry, M.J. (1983) {\it{Positive mass theorems for black
holes}}, Commun.Math.Phys. {\bf{88}}, 295-308

\bibitem[G80] {g80} Gibbons, G.W. (1980) {\it{Non-Existence Of Equilibrium Configurations Of Charged
Black Holes}}, Proc. Roy. Soc. Lond.A {\bf{372}}, 535-538.

\bibitem[G82] {g82} Gibbons, G.W. (1982) {\it{Antigravitating Black Hole Solitons With Scalar Hair In
N=4 Supergravity}}, Nucl. Phys.B {\bf{207}}, 337-349.

\bibitem[G93] {g93} Gibbons, G.W. (1993) {\it{No glory in cosmic string theory}}, Phys.Lett.{\bf{B308}}, 237-239

\bibitem[G02] {g02} Gibbons, G.W. (2002) {\it{The maximum tension principle in general relativity}},
Found. Phys. {\bf{32}}, 1891-1901 [arXiv:hep-th/0210109].

\bibitem[HR] {hr} Herdeiro, C.A.R. and Rebelo, C. (2008) {\it{On the interaction between two Kerr
black holes}}, JHEP {\bf{0810}}  017, 20 pp. [arXiv:0808.3941 [gr-qc]].

\bibitem[HRZ] {hrzc} Herdeiro, C.A.R., Rebelo, C., Zilhao, M., and Costa, M.S.
(2008) {\it{A Double Myers-Perry Black Hole in Five Dimensions}},
JHEP {\bf{0807}} 009, 24 pp. [arXiv:0805.1206 [hep-th]].

\bibitem[IK]{Israel} Israel, W. and   Khan, K.A. (1964)
{\it Collinear particles and Bondi dipoles in general relativity}
Il Nuovo Cimento {\bf 33}  1955-1965

\bibitem[KK] {kk} Kenneth, O.  and Klich, I. (2006) {\it{Opposites Attract - A Theorem About
The Casimir Force}}, Phys. Rev. Lett. {\bf{97}} 160401, 4 pp. [arXiv:0601011 [quant-ph]].

\bibitem[LMP] {lmp} Lu, L., Mei, J.,  and Pope, C.N. (2008) {\it{New Charged Black Holes in Five Dimensions}},
[arXiv:0806.2204 [hep-th]].

\bibitem[M] {Man}Manko V. S. (2007)  {\it  Double-Reissner-Nordstrom solution and the
interaction force between two spherical charged
masses in general relativity}, Phys Rev {\bf 76} 124032, 6 pp. [arXiv:0710.2158 [gr-qc]]

\bibitem[MP] {mp} Myers, R.C. and Perry, M.J. (1986) {\it{Black
Holes in Higher Dimensional Space-Times}}, Ann.Phys {\bf{172}},
304-347.

\bibitem[NH] {nh} Neugebauer, G. and Hennig, J. (2009) {\it{Non-existence of stationary two-black-hole
configurations}}, [arXiv:0905.4179 [gr-qc]]

\bibitem[SY] {sy} Schoen, R.M. and Yau, S.-T. (1981) {\it{Proof of the
positive mass theorem II}}, Commun.Math.Phys. {\bf{79}}, 231-260

\bibitem[SC] {s} Schoen, R.M., in preparation

\bibitem[S]{syn} Synge, J.L. (1969) {\it{Relativity: The General Theory}}, North-Holland, Amsterdam

\bibitem[T]{Taubes} Taubes, C. (1982) {\it The Existence
of a non-minimal solution to the $SU(2)$ Yang-Mills-Higgs Equations
on ${\mathbb R}^3$ : Part I} Commun. Math Phys {\bf 86} 257-298
{\it The Existence
of a non-minimal solution to the $SU(2)$ Yang-Mills-Higgs Equations
on ${\mathbb R}^3$ : Part II} Commun. Math Phys {\bf 86} 299-320

\bibitem[WA] {w} Wald, R. (1972) {\it{Gravitational Spin Interaction}}, Phys. Rev. {\bf{D 6}},
406-413.

\bibitem[WE]{w94} Weinstein, G. (1994)
 {\it{On the Force Between Rotating Coaxial Black Holes}},
Trans.Amer.Math.Soc. {\bf{343}}, 899-906.

\bibitem[WH] {wh} Whittaker, E.T. (1911) {\it History of the Theories
of Aether and Electricity from the Age of Descartes to the Close
of the Nineteenth Century} Longmans, Green and Co.
\end{thebibliography}
\end{document}